\documentstyle[12pt]{article}

\begin{document}

\title{Relativistic P\" oschl-Teller and Rosen-Morse problems}

\author{Ion I. Cot\u aescu \\{\it The West University of Timi\c soara,}\\
        {\it V. P\^ arvan Ave. 4, RO-1900 Timi\c soara, Romania}}

\date{\today}

\maketitle

\begin{abstract}
It is shown that  the new family of geometric models of the relativistic 
oscillator \cite{COTA}, which generalize the anti-de Sitter model, leads 
to  relativistic P\" oschl-Teller or Rosen-Morse problems. 
\end{abstract}
\

One of the simplest (1+1) geometric models is that of the  (classical or 
quantum) relativistic harmonic oscillator (RHO). Based on  phenomenological 
\cite{P1} and group theoretical \cite{A1,A2} arguments, this has been  
defined as a free system on the  anti-de Sitter  static background. There 
exists a (3+1) anti-de Sitter static metric \cite{P1} which reproduces the 
classical equations of motion of the  non-relativistic izotropic harmonic 
oscillator. This metric can be  restricted to  a (1+1) metric which  gives the 
classical equation of the one-dimensional non-relativistic harmonic oscillator 
(NRHO). Moreover, the corresponding quantum model has an equidistant discrete 
energy spectrum with a ground state energy larger, but approaching $\omega/2$ 
in the non-relativistic limit (in natural units $\hbar =c=1$) \cite{M}. 

In a previous article \cite{COTA} we have generalized this model to the family 
of (1+1) models depending on a real parameter $\lambda$ which has the metrics 
given by
\begin{equation}\label{(m)}
ds^{2}=g_{00}dt^{2}+g_{11}dx^{2}=
\frac{1+(1+\lambda) \omega^{2}x^{2}}{1+\lambda \omega^{2} x^{2}}dt^{2}-  
\frac{1+(1+\lambda) \omega^{2}x^{2}}{(1+\lambda \omega^{2} x^{2})^{2}}dx^{2},
\end{equation} 
where $\omega$ is the  frequency. This parametrization has been 
defined in order to obtain the exact anti-de Sitter metric  for $\lambda = -1$. 
Therein we have studied the scalar field $\phi$ of the mass $m$, defined on a 
suitable space domain, $D$, and minimally coupled with the gravitational field 
\cite{B1} given by (\ref{(m)}). Since the  energy is conserved, the 
Klein-Gordon equation, $(\Box + m^{2})\phi =0$, admits the set of fundamental 
solutions (of positive and negative frequency)  
\begin{equation}\label{(sol)}
\phi_{E}^{(+)}=\frac{1}{\sqrt{2E}}e^{-iEt}U_{E}(x), \quad 
\phi^{(-)}=(\phi^{(+)})^{*},
\end{equation}
which must be orthogonal with respect to the relativistic scalar product 
\cite{B1}. This reduces to the following scalar product of the wave 
functions $U$
\begin{equation}\label{(psc1)}
(U,U')=\int_{D}dx\mu(x)U^{*}(x)U'(x),
\end{equation}
where
\begin{equation}\label{(10)}
\mu(x)=\sqrt{-g(x)}g^{00}(x) = \frac{1}{\sqrt{1+\lambda\omega^{2} x^{2}}}, 
\quad g=det(g_{\mu\nu}).
\end{equation}
By solving the Klein-Gordon   equations  we have shown that  
the models with $\lambda>0$ have mixed energy spectra, with a finite discrete 
sequence and a continuous part, while for $\lambda\le 0$ these spectra are 
countable \cite{COTA}. However, despite of their  different relativistic 
behavior, all these models have the same non-relativistic limit, namely the 
NRHO of the frequency $\omega$. For this reason we shall use the name of 
relativistic oscillators (RO) for all the models with $\lambda \not= -1$, 
understanding that the RHO is only that of the anti-de Sitter metric. 

In general,  any (1+1)  static background admits a special natural frame  in 
which the metric is a conformal transformation of the Minkowski flat metric. 
This  new frame can be obtained by changing the space coordinate   
\begin{equation}\label{(100)}
x\to \hat x = \int dx\mu(x) + const.
\end{equation}
Then  
\begin{equation}
\hat g_{00}(\hat x)=-\hat g_{11}(\hat x)=\sqrt{-\hat g(\hat x)}, 
\quad \hat\mu(\hat x)=1 
\end{equation}
and, therefore, the scalar product (\ref{(psc1)}) becomes the usual one. 

Our aim is to show that, in this frame, the Klein-Gordon 
equation  of our RO will involve relativistic symmetric P\" oschl-Teller (PT) 
potentials for $\lambda<0$ and symmetric Rosen-Morse (RM) potentials for 
$\lambda > 0$ \cite{PT}. When $\lambda=0$ the relativistic potential will be 
proportional with that of the NRHO. 

Let us first consider the case of $\lambda<0$ and  denote 
\begin{equation}\label{(par2)}
\lambda=-\epsilon^{2}, \qquad \hat\omega =\epsilon \omega, \qquad 
\epsilon\ge 0.
\end{equation} 
Then, by changing the space coordinate according to (\ref{(100)}) and 
(\ref{(10)})  we obtain 
\begin{equation}
\hat x=\frac{1}{\hat\omega}\arcsin\hat\omega x
\end{equation}
and  the new form of the line element 
\begin{equation}
ds^{2}=\left(1+\frac{1}{\epsilon^2}\tan^{2}\hat\omega \hat x\right)(dt^{2}-
d\hat x^{2}).
\end{equation} 
The functions $U$ are defined on  $D =(-\pi/2\hat\omega, \pi/2\hat\omega)$ 
because of the singularities of the metric which determine the event horizon. 
The Klein-Gordon equation,
\begin{equation}\label{(kg2)}
\left(-\frac{d^2}{d\hat x^{2}} +\frac{m^2}{\epsilon^2}\tan^{2}\hat\omega\hat x
\right)U(\hat x)=({E}^{2}-m^{2})U(\hat x),
\end{equation}
has a countable discrete energy spectrum   \cite{COTA}.
Therefore, the energy eigenfunctions  must be square integrable with respect 
to the usual scalar product in the coordinate $\hat x$. These have the form 
\begin{equation}\label{(U2)}
U_{n}(\hat x)=N_{n_{s},s} \cos^{k}\hat\omega\hat x 
\sin^{s} \hat\omega\hat x F(-n_{s},k+s+n_{s},
s+\frac{1}{2},\sin^{2} \hat\omega \hat x),
\end{equation}
where  $N_{n_{s},s}$ is the normalization factor, and  
\begin{equation}\label{(p)}
 k=\frac{1}{2}\left[1+\sqrt{1+ 4\frac{m^{2}}{\epsilon^{2}\hat\omega^{2}}}
  \right] > 1,
\end{equation}
is the positive solution of the equation
\begin{equation}\label{(kk)}
k(k-1)=\frac{m^2}{\epsilon^{2}\hat\omega^2}.
\end{equation}
The quantum numbers $n_{s}=0,1,2...$ and $s=0, 1$  can be embedded into the 
main quantum number $n=2n_{s}+s$, which take  even values if $s=0$ and odd 
values for $s=1$.  The energy levels are given by 
\begin{equation}\label{(el1)}
{E_{n}}^{2}=m^{2}+\hat\omega^{2}[2k(n+\frac{1}{2})+n^{2}], \quad 
n=0,1,2... .
\end{equation}  
if $\epsilon\not=1$, and by
\begin{equation}\label{(el2)}
E_{n}=\hat\omega(k+n)
\end{equation}
in  the tha case of the RHO \cite{M}, when $\epsilon=1$. 
According to (\ref{(kk)}), the second term of the left-hand side  
of (\ref{(kg2)}) can be written as  
\begin{equation}
V_{PT}(\hat x)=k(k-1)\hat\omega^2\tan^{2}\hat\omega \hat x.
\end{equation}
This will be called the relativistic (symmetric) PT 
potential since the solutions (\ref{(U2)})  coincide with those 
given by the Schr\" odinger equation with the non-relativistic PT potential 
$V_{PT}/2m$.  Hence, for $\epsilon >0$ our RO are systems of relativistic 
massive scalar particles confined to wells, as in the non-relativistic case, 
but having new  energy spectra and another parametrization which depends 
on $m$, $\omega$ and $\epsilon$.  We note that our new parameter 
$\epsilon$  allows to choose  the desired well width, $\pi/\epsilon\omega$, 
when the frequency $\omega$ is fixed. Therefore, this will be a supplementary 
fit parameter in the problems of geometric confinement.

For $\lambda>0$ we change the significance of $\epsilon$ and we put 
\begin{equation}\label{(par4)}
\lambda=\epsilon^{2}, \qquad \hat\omega =\epsilon \omega, \qquad 
\epsilon\ge 0,
\end{equation} 
so that, according to (\ref{(100)}) and (\ref{(10)}), the change of the 
space coordinate will be given by    
\begin{equation}
x=\frac{1}{\hat\omega}\sinh\hat\omega \hat x .
\end{equation}
Now, the  line element is 
\begin{equation}
ds^{2}=\left(1+\frac{1}{\epsilon^2}\tanh^{2}\hat\omega \hat x\right)(dt^{2}-
d\hat x^{2})
\end{equation} 
and  $D =(-\infty, \infty)$. 
The Klein-Gordon equation 
\begin{equation}\label{(kg2)}
\left(-\frac{d^2}{d\hat x^{2}} +\frac{m^2}{\epsilon^2}\tanh^{2}\hat\omega\hat x
\right)U(\hat x)=({E}^{2}-m^{2})U(\hat x)
\end{equation}
has a mixed energy spectrum \cite{COTA}. The  square integrable energy 
eigenfunctions of the finite discrete spectrum are  
\begin{equation}\label{(U3)}
U_{n}(\hat x)=N_{n_{s},s} \cosh^{-k'}\hat\omega\hat x 
\sinh^{s} \hat\omega\hat x F(-n_{s},-k'+s+n_{s},
s+\frac{1}{2},-\sinh^{2} \hat\omega \hat x),
\end{equation}
where now    
\begin{equation}\label{(p)}
 k'=\frac{1}{2}\left[-1+\sqrt{1+ 4\frac{m^{2}}{\epsilon^{2}\hat\omega^{2}}}
  \right] 
\end{equation}
is the positive solution of the equation
\begin{equation}\label{(jj)}
k'(k'+1)=\frac{m^2}{\epsilon^{2}\hat\omega^2}.
\end{equation}
The quantum number $n=2n_{s}+s$ can take the values $n=0,1,...n_{max}<k'$.  
One can verify that the discrete spectrum is included in the domain 
$[m, m\sqrt{1+1/\epsilon^{2}})$ since the energy levels are given by 
\begin{equation}\label{(el1)}
{E_{n}}^{2}=m^{2}+\hat\omega^{2}[2k'(n+\frac{1}{2})-n^{2}], \quad 
n=0,1,2... n_{max}
\end{equation}  
The continuous spectrum is $[m\sqrt{1+1/\epsilon^{2}}, \infty)$ while the 
corresponding generalized energy eigenfunction are    
\begin{equation}\label{(U4)}
U_{\nu}(\hat x)=N_{\nu} \cosh^{-k'}\hat\omega\hat x 
\sinh^{s} \hat\omega\hat x F(-k'+s+i\nu,-k'+s-i\nu,
s+\frac{1}{2},-\sinh^{2} \hat\omega \hat x),
\end{equation}
where
\begin{equation}
\nu(E)= \frac{1}{2\hat\omega}\sqrt{E^{2}-m^{2}\left(1+\frac{1}{\epsilon^2}
\right)} \in [0, \infty) .
\end{equation} 
As in the previous case, we shall say that  
\begin{equation}
V_{RM}(\hat x)=k'(k'+1)\hat\omega^2\tanh^{2}\hat\omega \hat x
\end{equation}
is the relativistic  (symmetric) RM  potential. The solutions (\ref{(U3)}) 
and (\ref{(U4)}) are the same as  those given by the non-relativistic RM 
potential $V_{RM}/2m$. Of course, the  relativistic energy spectra  differ 
from the non-relativistic ones. We note  that the number of the discrete energy 
levels is determined by the values of $m/\omega$ (i.e. $mc^{2}/\hbar\omega$ in 
usual units) and  $\epsilon$ defined according to (\ref{(par4)}). Moreover, for 
a fixed $m$, the domains of the discrete and continuous spectra are given by 
$\epsilon$ only.

We have shown \cite{COTA} that our family of RO is continuous in 
$\lambda=0$. This means that the limits for $\epsilon \to 0$ of the PT and RM 
systems must coincide.  Indeed, then we have $\hat x\to x$ and $k \to 
\infty$, $k'\to \infty$ but
\begin{equation}
\lim_{\epsilon \to 0}\epsilon^{2}k=
\lim_{\epsilon \to 0}\epsilon^{2}k'=\frac{m}{\omega}.
\end{equation}
Therefore, we can verify that, in this limit, we obtain the relativistic 
potential  
\begin{equation}
V(x)=\lim_{\epsilon\to 0} V_{PT}(\hat x)=  
\lim_{\epsilon\to 0} V_{RM}(\hat x)=m^{2}\omega^{2}x^{2}  
\end{equation}
which gives the same energy eigenfunctions as those of the NRHO and the levels
\begin{equation}
E^{2}_{n}=m^{2}+2m\omega(n+\frac{1}{2}) . 
\end{equation}
Obviously, the NRHO potential is $V/2m$.

The conclusion is that our family of  metrics (\ref{(m)}), which depend on
the parameter $\lambda$, generates  relativistic PT or RM problems, in the 
special frames $(t, \hat x)$. The PT and RM potentials have similar forms being  
proportional with $m^{2}/|\lambda|$. On the other hand, it is interesting that 
this parameter (or the parameter $\epsilon$ related to it) has not a direct 
non-relativistic equivalent since, in this limit, all the RO become NRHO and, 
consequently, the terms involving $\lambda$ disappear \cite{COTA}. However, its 
physical significance results from the  analysis of the relativistic effects, 
as we have seen from the previous discussion concerning the behavior of the  
relativistic PT or RM systems.

\end{document}